\documentclass[aps,prl,twocolumn,showpacs]{revtex4}

\usepackage{graphicx}

\begin{document}
\def\ba{BaFe$_{2}$As$_{2}$}
\def\co{Ba(Fe$_{0.93}$Co$_{0.07}$)$_{2}$As$_{2}$}
\def\baco{Ba(Fe$_{1-x}$Co$_{x}$)$_{2}$As$_{2}$}
\def\htc{high-$T_{c}$ cuprates}
\def\htsc{high temperature superconductors}
\def\jc{$J_{c}$}
\def\etal{{\it et al.}}


\title{Enhancement of Critical Current Densities in Co-Doped BaFe$_{2}$As$_{2}$ with Columnar Defects Introduced by Heavy-Ion Irradiation}


\author{Y.~Nakajima,$^{1,2}$ Y.~Tsuchiya,$^{1}$ T.~Taen,$^{1}$ T.~Tamegai,$^{1,2}$ S.~Okayasu,$^{3}$ and M.~Sasase$^{4}$}
\affiliation{$^{1}$Department of Applied Physics, The University of Tokyo, Hongo, Bunkyo-ku, Tokyo 113-8656, Japan}
\affiliation{$^{2}$JST, Transformative Research-Project on Iron Pnictides (TRIP), 7-3-1 Hongo, Bunkyo-ku, Tokyo 113-8656, Japan}
\affiliation{$^{3}$Advanced Science Research Center, Japan Atomic Energy Agency, Tokai, Ibaraki 319-1195, Japan}
\affiliation{$^{4}$The Wakasa-wan Energy Research Center, Research and Development Group, 64-52-1 Nagatani, Tsuruga, Fukui 914-0192, Japan}


\date{\today}

\begin{abstract}
We report the first realization of columnar defects in Co-doped BaFe$_{2}$As$_{2}$ single crystals by heavy-iron irradiation. The columnar defects are confirmed by transmission electron microscopy and their density is about 40 $\%$ of the irradiation dose. Magneto-optical imaging and bulk magnetization measurements reveal that the critical current density is strongly enhanced in the irradiated region. We also find that vortex creep rates are strongly suppressed by the columnar defects. We compare the effect of heavy-ion irradiation into Co-doped {\ba} and cuprate superconductors.
\end{abstract}


\maketitle



A limitation on the technological advances of high-temperature superconductors comes from the intrinsically low critical current density {\jc}.  Recently discovered iron-based superconductors \cite{kamih08} also face the same problem. While the transition temperature $T_{c}$ is increased up to $\sim$ 55 K in rare-earth-based iron-oxyarsenides within a short period of time \cite{ren08}, {\jc} at low temperatures is still low \cite{prozo08,yamam09,nakaj09}. Although the transition temperature of iron-pnictide superconductors is still lower than cuprate superconductors, introduction of pinning centers can enhance the critical current density and make this system more attractive for practical applications. It is well known that a most efficient way to improve the critical current density is to pin vortices with columnar defects created by swift particle irradiation. In {\htsc}, columnar defects enhances {\jc} dramatically \cite{civale91,konczy91}. 

Intermetallic iron-arsenides {\baco} with $T_{c}\sim$ 24 K is readily available in large single crystalline form \cite{sefat08} and its {\jc} reashes 10$^{6}$ A/cm$^{2}$ at $T$ = 2 K, which is potentially attractive for technological applications \cite{prozo08,yamam09,nakaj09}. We expect that {\jc} in {\baco} could be enhanced by introducing columnar defects that can pin vortices. However, it is well known that irradiation-induced defects strongly depend on various parameters such as ion energy, stopping power of incident ions, thermal conductivity and perfection of the target crystal, etc. \cite{zhu93} Since there are so many influencing factors, it is still an open question whether irradiation damage can be introduced in iron-arsenide superconductors. In this paper, we report the first attempt to create columnar defects by heavy-iron irradiation into Co-doped BaFe$_{2}$As$_{2}$ single crystals. Columnar tracks with diameters of $\sim$ 2-5 nm and about 40\% of nominal ion dose are clearly seen in scanning transmission electron microscopy (STEM) images. Magneto-optical images and bulk magnetization measurements reveal a strong enhancement of {\jc} in the irradiated region. Columnar defects also suppress the relaxation of magnetization consistent with the enhancement of pinning capability of vortices. Effect of columnar defects on the vortex dynamics is compared between Co-doped {\ba} and high temperature superconductors.


Single crystalline samples of {\co} were grown by FeAs/CoAs self-flux method \cite{nakaj09}.  Co concentration was determined by EDX measurements. 200 MeV Au ions were irradiated into {\co} along $c-$axis using the TANDEM accelerator in JAEA to create columnar defects. In order to maximize the effect of irradiation, crystals with thicknesses $\sim$ 10 $\mu$m were irradiated considering the penetration depth of 200 MeV Au ions into {\co}, 8.5 $\mu$m. Field-equivalent defect densities, matching field, is $B_{\phi} =$ 20 kG. Plan view and cross-sectional observations of the irradiated {\co} were performed with a high resolution and scanning TEM (JEOL, JEM-3000F). Magnetization was measured by a commercial SQUID magnetometer (MPMS-XL5, Quantum Design). Magneto-optical images were obtained by using the local-field-dependent Faraday effect in the in-plane magnetized garnet indicator film employing a differential method \cite{soibe00,yasug02}. 

\begin{figure}[t]
\begin{center}
\includegraphics[width=7cm]{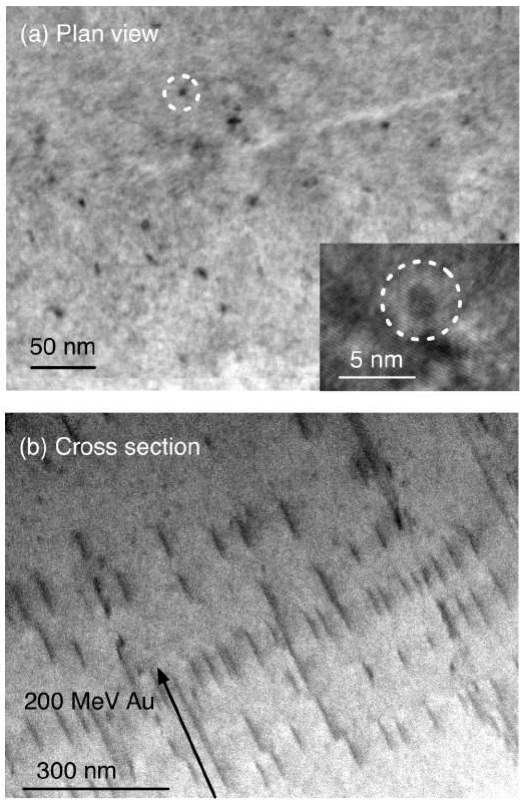}
\caption{STEM observation of an irradiated {\co}. (a) Plan view. Dotted circle indicates the columnar defect with average density of $4.2\pm1.2\times 10^{10}$ cm$^{-2}$, which corresponds to about 40\% of the expected value. Inset: High resolution TEM image (different region). (b) Cross section. The arrow indicates the direction of 200 MeV Au irradiation, which corresponds to $c-$axis.}
\label{FIG1}
\end{center}
\end{figure}

\begin{figure}[t]
\begin{center}
\includegraphics[width=8cm]{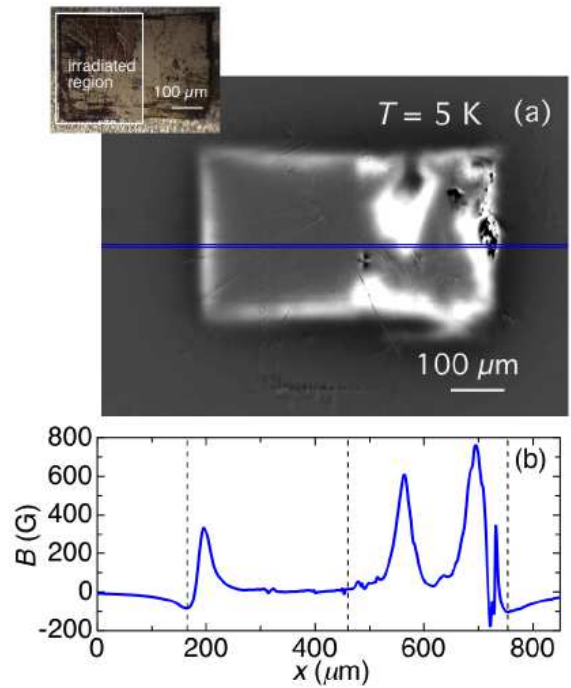}
\caption{(color online) (a) Magneto-optical image of the remanent state in {\co} at 5 K. The inset shows the optical image of {\co}. The area surrounded by the rectangle is the region irradiated by 200 MeV Au ions. (b) Line profile along a line in Fig. 2(a). Broken lines show locations of sample edges and the irradiation boundary.}
\label{FIG2}
\end{center}
\end{figure}

\begin{figure}[t]
\begin{center}
\includegraphics[width=8cm]{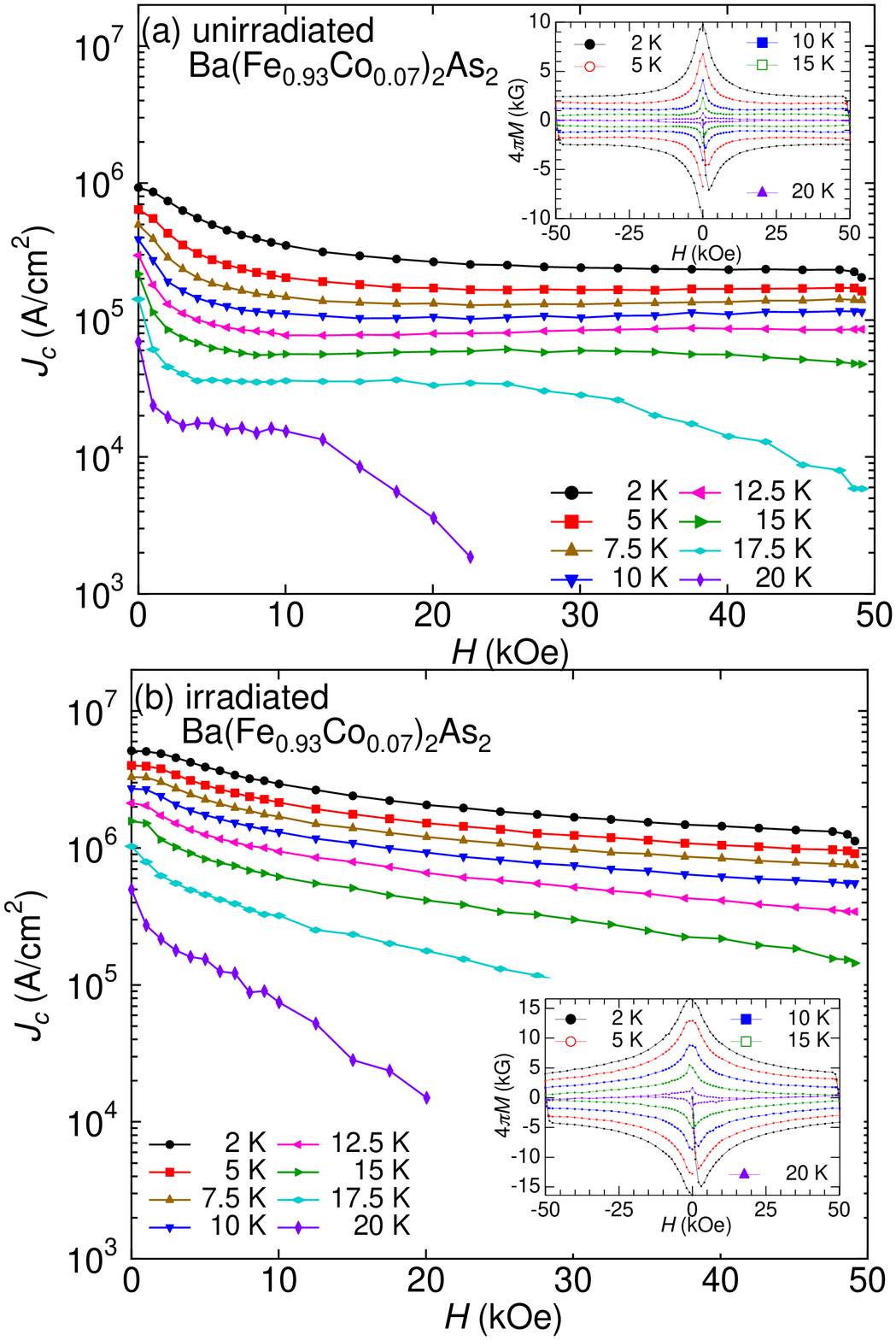}
\caption{(color online) Field dependence of critical current density calculated by Eq. (2) at several temperatures in (a) the unirradiated ($164\times309\times30$ $\mu$m$^{3}$) and (b) the irradiated ($450\times584\times30$ $\mu$m$^{3}$) {\co}.  Inset: Field dependence of the  magnetization at several temperatures in the unirradiated and irradiated {\co}.  }
\label{FIG3}
\end{center}
\end{figure}

\begin{figure}[t]
\begin{center}
\includegraphics[width=8cm]{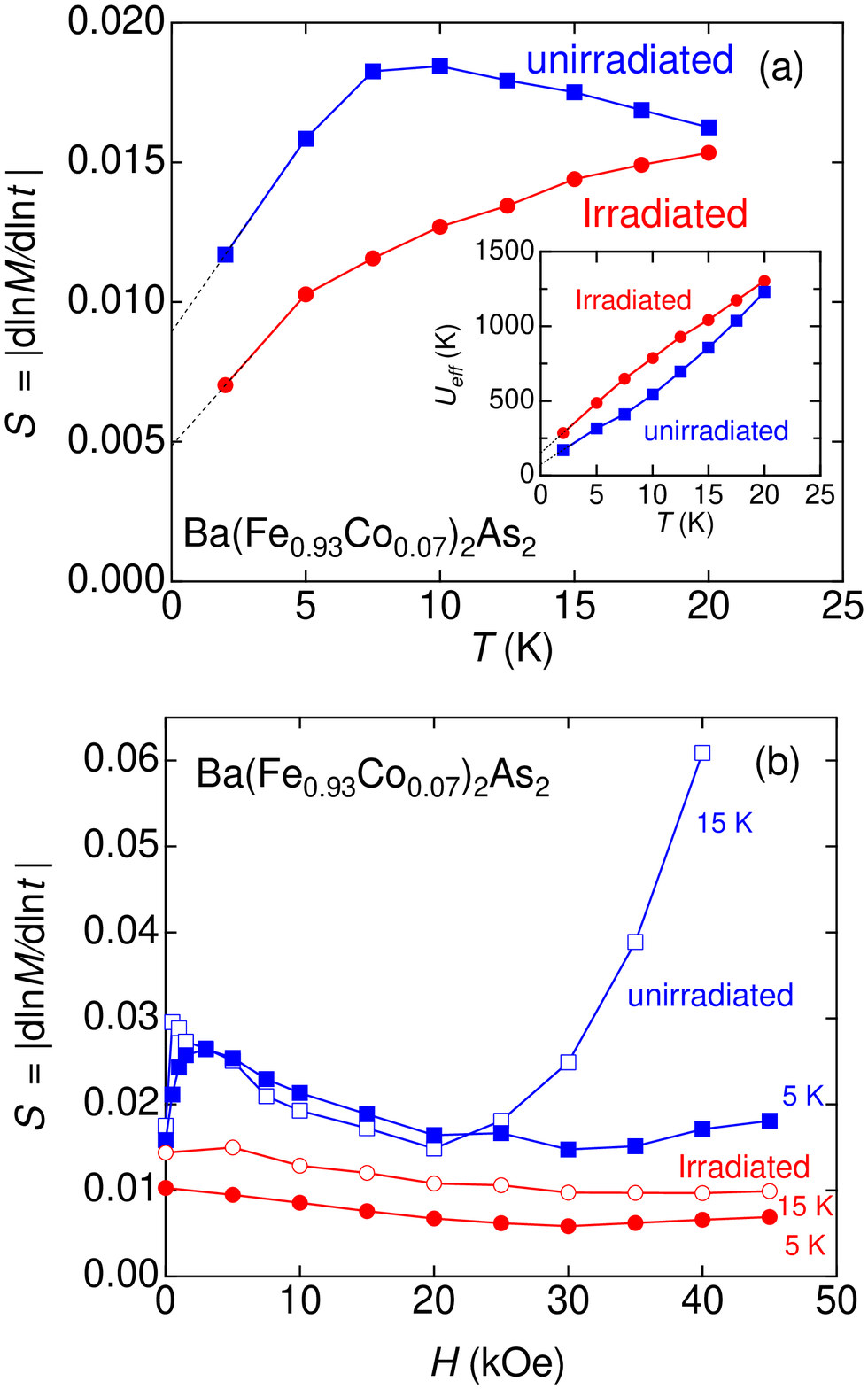}
\caption{(color online) (a) Temperature dependence of normalized relaxation rate $S=|d\ln M/d\ln t|$ in the unirradiated and irradiated {\co} in the remanent state. Dotted lines are linear extrapolations down to $T=0$ K. Inset: Temperature dependence of effective pinning energy $U_{eff}=T/S$ in the  unirradiated and the irradiated {\co}. Dotted lines are linear extrapolations down to $T=0$ K. (b) Field dependence of $S$ in the unirradiated and irradiated {\co} at 5 K and 15 K.  }
\label{FIG4}
\end{center}
\end{figure}


Figure 1(a) shows a plan view of the irradiated {\co} along $c-$axis, where defects are marked by dotted circles. The density of defects extracted from the average of ten different regions is 4.2$\pm$1.2$\times$10$^{10}$ cm$^{-2}$, which is about 40\% of the expected value. High resolution TEM observation shown in the inset of Fig. 1(a) reveals that the size of defects is $\sim$ 2-5 nm with a large fluctuation, and the lattice
image is sustained with some displacements of atoms. Such characteristics of heavy-ion induced defects makes a good contrast to amorphous defects with diameters $\sim$ 10 nm created in high temperature cuprate superconductors \cite{zhu93}. The morphology of defects along the projectile is clearly columnar as identified in the cross sectional image shown in Fig. 1(b). However, columnar defects are discontinuous with various lengths ranging 30-240 nm. We also note that our preliminary measurements on {\co} irradiated by 200 MeV Ni ions give reduced effect on its electromagnetic properties. All these experimental facts suggests that 200 MeV Au irradiation is marginal to create continuous columnar defects in {\co}. Creation of columnar defects are controlled by various factors, such as mass and energy of heavy ions, thermal conductivity and carrier density of the system. It would be of great interest to compare the defect characteristics created by lighter and heavier ions.

Enhancements of vortex pinning introduced by the irradiation are directly visualized by magneto-optical imaging. The inset of Fig. 2(a) shows an optical image of {\co}.  For comparison, we covered a half of the crystal by Au foil with a thickness of 100 $\mu$m and the rest half is irradiated. The left area inside the rectangle is irradiated region. We note that $T_{c}$ of the irradiated {\co} sample is 24 K and is not affected by the introduction of the columnar damage. Main panel of Fig. 2(a) shows a magneto-optical image in the remanent state of {\co} at 5 K after increasing field up to 800 Oe for 5 s and reducing it to zero. We find prominent difference of vortex penetrations between the irradiated and unirradiated regions. In the unirradiated region, vortices penetrate deep into the sample in addition to the penetration from a defect on the top edge. By contrast, the penetration in the irradiated region is much reduced than those in the unirradiated region, which indicates that defects introduced by the irradiation enhance the vortex pinning in the irradiated region. In fact, critical current density in the irradiated region is strongly enhanced. Line profile obtained from Fig. 2(a) is plotted in Fig. 2(b). The critical current density can be roughly estimated using the equation,
\begin{equation}
	J_{c}=\frac{c}{4d}\frac{H_{ex}}{\cosh^{-1}(w/(w-2p))},
\end{equation}
where $H_{ex}$ is external field, $w$ is the sample width, $d$ is the sample thickness, and $p$ is the penetration with measured from the sample edge \cite{brand93}. From this equation with the width of the irradiated region $w$ = 295 $\mu$m, $d$ = 10 $\mu$m, and $p=59$ $\mu$m, we can estimate roughly $J_{c}$ as $\sim 1.8\times 10^{6}$ A/cm$^{2}$ in the irradiated region, which is much larger than that in the unirradiated sample \cite{prozo08,yamam09,nakaj09} and demonstrates the effectiveness of columnar defects for vortex pinning in Co-doped {\ba}.

We can also confirm the enhancement of flux pinning by the bulk magnetization measurements. To evaluate the magnetization precisely, we separated the irradiated region by carefully cutting the half-irradiated sample. Insets of Figs. 3(a) and (b) show the magnetization curves at several temperatures in the unirradiated and irradiated samples, respectively. In the unirradiated sample, pronounced fish-tail effect is observed at higher temperatures and the peak at zero field is very sharp \cite{prozo08,yamam09,nakaj09}. By contrast, the magnetization in the irradiated sample is enhanced in spite of much smaller size and the fish-tail effect is absent. The peak at zero field is broadened due to the drastic increase of low-field irreversible magnetization.  We note that in {\htc} a peak near $\sim B_{\phi}/3$ in magnetization is observed \cite{sato97,itaka01}. In this system, however, the prominent peak of irreversible magnetization at $\sim B_{\phi}/3$ may be absent or very small.

The bulk critical current density $J_{c}$ is derived from the Bean model,
\begin{equation}
	J_{c}=20\frac{\Delta M}{a(1-b/3a)},
\end{equation}
where $\Delta M$ is $M_{down}-M_{up}$, $M_{up}$ and $M_{down}$ are the magnetization when sweeping fields up and down, respectively, and $a$ and $b$ are the sample widths $(a<b)$. Figures 3(a) and (b) shows the field dependence of $J_{c}$ in the unirradiated and irradiated samples, respectively. In the unirradiated sample, {\jc} is about $6.4\times10^{5}$ A/cm$^{2}$ at $T=$ 5 K under zero applied field. At high temperatures {\jc} changes non-monotonically with magnetic field reflecting the fish-tail effect in magnetization. In the irradiated sample, a drastic enhancement of {\jc} is observed in the whole temperature and field range. {\jc} in the irradiated sample at $T=5$ K under zero field reaches $4.0\times10^{6}$ A/cm$^{2}$. {\jc} in the irradiated sample decreases monotonically with increasing magnetic field. These results obtained from bulk magnetization measurements also indicate that heavy-ion irradiation introduces pinning centers for vortices.

To get insight into vortex dynamics, we investigate the relaxation of shielding current. Figure 4 shows the normalized relaxation rate $S$ defined by $S=|d\ln M/d\ln t |$ in the remanent state of the unirradiated and irradiated samples. In the unirradiated sample, $S$ shows a broad peak around $T\sim$ 8 K. This behavior is qualitatively similar to that reported in Ref. \cite{prozo08}. By contrast, $S$ in the irradiated sample is strongly suppressed at all temperatures, which demonstrates the effect of columnar defects on flux creep. With increasing temperature, $S$ in the irradiated sample increases monotonically and approaches to that in the unirradiated sample at temperatures close to $T_{c}$, which indicate that additional flux pining by columnar defects is reduced by thermal fluctuations. It should be noted that $S$ shows a steep increase close to zero field in the unirradiated sample. The peak of $S$ around 8 K in the unirradiated sample could be related to this low-field anomaly. Namely, as temperature is lowered, trapped field in the sample increases, inducing the increase of $S$ in average due to the intrinsic field dependence of $S$. In fact, the temperature dependence of $S$ in the irradiated sample, where no low-field increase is observed, does not show a maximum. We note that  in YBa$_{2}$Cu$_{3}$O$_{7-\delta}$,  temperature independent plateau in the normalized relaxation rate with values in the range $S=0.020-0.035$, similar to the present value, is observed and interpreted in terms of collective creep theory \cite{maloz90}.

To get further insight, we estimate the pinning energy $U_{eff}$ using the conventional flux creep relation
\begin{equation}
	M=M_{0}\left(1-\frac{T}{U_{eff}}\ln\left(t/t_{0}\right)\right ),
\end{equation}
where $t_{0}$ is the vortex-hopping attempt time. From this relation, effective pinning energy is obtained as $U_{eff}=T/S$. Inset of Fig. 4(a) shows the temperature dependence of $U_{eff}$ in the unirradiated and irradiated samples. $U_{eff}$ in both samples increases almost linearly with temperature, which is similar to that in oxy-pnictide superconductor SmFeAsO$_{0.9}$F$_{0.1}$ \cite{yang08}.  Enhancement of $U_{eff}$ by the columnar defects is prominent at intermediate temperatures and disappear at high temperatures. By extrapolating the curve of $U_{eff}$ down to $T=0$, we obtain $U_{eff}(0)\sim$ 75 K and $\sim$ 150 K in the unirradiated and irradiated samples, respectively. Columnar defects  enhance the value of $U_{eff}(0)$ by a factor of 2. It should be noted that $U_{eff}(0)$ is 40 K in SmFeAsO$_{0.9}$F$_{0.1}$ \cite{yang08} and 100 - 400 K in YBa$_{2}$Cu$_{3}$O$_{7-\delta}$ thin films \cite{wen95}.

Figure 4(b) shows the field dependence of $S$ at 5 K and 15 K in the unirradiated and irradiated samples. In the unirradiated sample at 5 K, $S$ is nearly field independent except for a drastic increase at low fiedls ($H\lesssim 3$ kOe), which could be related with the presence of Meissner hole in the remanent state \cite{vlask97}. At 15 K, $S$ exhibit a minimum at 20 kOe after showing a drastic increase at very low fields. This behavior at 15 K is very similar to that reported in Ref. [\onlinecite{prozo08}]. We note that strong field dependence at 15 K may be related to the fish-tail effect.

While in the framework of the classical creep model considering thermal activation of the vortices, $S$ should vanish at $T=0$ K, non-zero relaxation rate extrapolated to $T=0$ K due to quantum creep is observed \cite{blatt91}. At low temperatures, quantum creep is dominant, where $S$ is given by
\begin{equation}
	S\simeq \frac{e^{2}}{\hbar}\frac{\rho_{n}}{\xi}\left(\frac{J_{c}}{J_{0}}\right)^{\frac{1}{2}},
\end{equation}
where $\rho_{n}$ is the normal-state resistivity, $\xi$ is coherence length, and $J_{0}$ is depairing current density \cite{blatt91}. To estimate the quantum creep rate of {\baco}, we extrapolate $S$ linearly down to $T=0$ K. The extrapolated values to $T=0$ K are 0.009 and 0.005 in the unirradiated and irradiated samples, respectively.  $\rho_{n}$ is $\sim$ 130 $\mu\Omega$cm and $\xi_{ab}$ is obtained as $\sim 34$ \AA ~from $H_{c2}(0)\sim 280$ kOe \cite{nakaj09}, and $J_{0}=c\phi_{0}/12\sqrt{3}\pi^{2}\xi_{ab}\lambda_{ab}^{2}$. Therefore, we obtain $J_{0}\sim 7.4\times10^{7}$ A/cm$^{2}$ with $\lambda_{ab}\sim$ 2000 \AA~\cite{prozo09}. Using Eq. (3), we can obtain $S\sim 0.011$, which is quite consistent with the experimental results.

Finally, we comment the accommodation of vortices to columnar defects. In cuprate superconductors with columnar defects, vortices are localized onto columnar defects at low temperatures, where Bose-glass phase is expected \cite{nelson92}. In fact, distinct vortex dynamics related to the Bose-glass phase, for instance, a large peak of $S$ at $T\sim T_{c}/2$, is observed \cite{elbau96,thomp97}. However, in the irradiated {\co}, there is no acceleration of $S$ at intermediate temperatures. This could be due to discontinuous columnar defects as shown in Fig. 1(b), which reduces the localization of vortices onto columnar defects.

In summary, we have introduced columnar defects in Co-doped BaFe$_{2}$As$_{2}$ single crystals by 200 MeV Au ion irradiation. Formation of columnar defects are confirmed by STEM observation and their density corresponds to 40 \% of the expected value. Magneto-optical imaging and bulk magnetization measurements reveal that $J_{c}$ is strongly enhanced in the irradiated region. We also find that vortex creep rates are strongly suppressed by the columnar defects. We conclude that the columnar defects introduced by heavy-iron irradiation strongly enhance the flux pining in Co-doped BaFe$_{2}$As$_{2}$. 

We thank J. R. Thompson for valuable comments. This work is partly supported by a Grant-in-Aid for Scientific Reserch from MEXT.

\end{document}